\documentstyle[10pt,epsf]{article}

\newcommand{\beq}{\begin{equation}}
\newcommand{\eeq}{\end{equation}}

\newcommand{\eq}[1]{eq.(\ref{#1})}

\begin{document}

\title {
What do we actually know on the proton radius ?
}

\author {Savely G. Karshenboim\\
\medskip
{}\\
D. I.  Mendeleev Institute for Metrology (VNIIM),\\  St. Petersburg 198005,
Russia\thanks{E-mail: sgk@onti.vniim.spb.su}\\
{}\\
and\\
{}\\
Max-Planck-Institut f\"ur
Quantenoptik, 85748 Garching, Germany\thanks{Temporary address}
\thanks{E-mail: sek@mpq.mpg.de}
}


\maketitle

\begin{abstract}

The work is devoted to a consideration of the different determinations
of the proton charge radius. It is demonstrated that the results
from the elastic electron-proton scattering have to be of a higher
uncertainty. A review of the hydrogen Lamb shift measurements and the radius
determination from them is also presented.

\end{abstract}

\section{Introduction}

The proton is the lightest and simplest stable hadronic system.
Investigation of its structure is quite important. The charge
radius determined by the charge distribution inside the proton is one of
the universal
fundamental physical constants, because it is important for a number of very
different physical problems.  Some recent precise results are collected in
Table \ref{t1} and also presented in Fig. 1.
The proton charge radius is defined here as $R_p=\sqrt{\langle r^2
\rangle}$.
One can see that the values are obtained from different branches of physics.
We give first a short description of all well known and most `popular'
results summarized there\footnote{Of course, the Lamb shift result 
mentioned there is too fresh to be `popular' but it is going to be.}.

\begin{table}
\begin{center}
\begin{tabular}{||c|c|c||}
\hline\hline
&&\\[-1ex]
Value         & Reference  & Method \\  [1ex]
\hline
\hline
&&\\[-1ex]
0.809(11)~{\rm fm}  & Stanford, 1963  & scattering experiment \& empirical fitting
\\ [1ex]
0.862(12)~{\rm fm} & Mainz, 1980  &  scattering experiment \& empirical fitting
\\ [1ex]
0.64(8)~{\rm fm} & Draper, Woloshin, Liu, 1990 & lattice QCD in
chiral limit \\ [1ex]
0.88(3)~{\rm fm} & Leinweber, Cohen, 1993 &
lattice QCD \&
chiral perturbation theory \\ [1ex]
0.847(9)~{\rm fm} &  Mainz, 1996   & dispersion relation fitting     \\ [1ex]
0.890(14)~{\rm fm} &  Garching, 1997   & hydrogen Lamb shift measurements \\[1ex]
\hline\hline
\end{tabular}
\vspace{5mm}
\caption{Proton charge radius \label{t1}}
\end{center}
\end{table}

\begin{figure}[h]
\epsfxsize=9cm
\centerline{\epsfbox{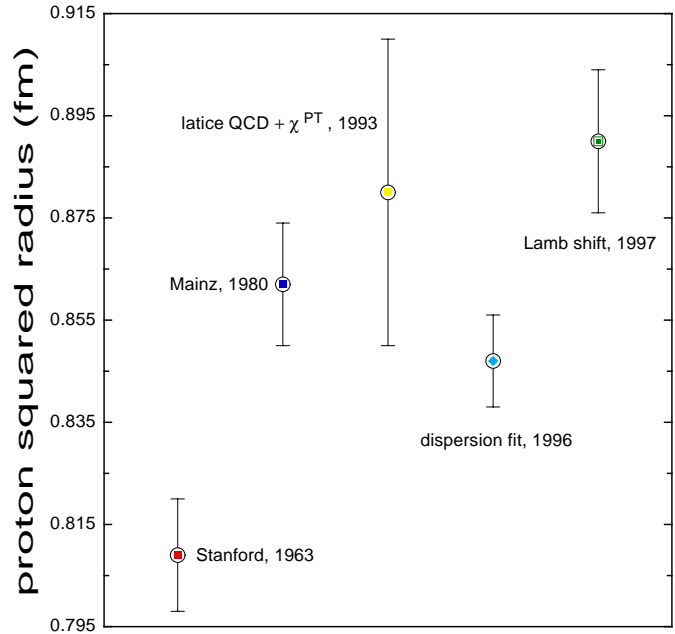}}
\caption{\label{f1} The proton charge radius as it is usually known}
\end{figure}

The references and methods used to determine the proton radius are:
\begin{itemize}
\item In paper \cite{Hand}
results on electron-proton  elastic scattering were obtained and
fitted using a simple empirical formula. The experiment has been done by
means of the accelerators and particle physics. Theoretically
they are based on the particle physics phenomenology and the quantum
electrodynamics theory for corrections.
\item  In work \cite{Simon} new results for $e-p$ were presented. The
radius was found by an extrapolation to zero transfer momentum.
\item In paper \cite{Draper} (see also Ref. \cite{Leinweber91})
a lattice calculation was performed within
the chiral limit, where the pion is massless ($m_\pi=0$). The result for the  radius was found by
fitting of the form factor computed there. The lattice
calculation are result of the quantum field theory without
using of the perturbative expansion in the Euclidian space. As far as the
space is not continuous there are a lot of problems there due to a formal
description of relations between the physics in the infinite continuous
Minkowsky space and in the discrete finite Euclidian one.
\item In article \cite{Leinweber93} a chiral perturbation correction
to previous calculations was presented. The chiral perturbation theory
is another branch of the quantum field theory where the smallness of
the masses of the $u$ and $d$ quarks is used.
\item In Ref. \cite{Mergell} a many parameters fit was performed. It
included transferred momentum within a wide range. At least two
new experimental subfields are involved into this approach. First the wider range of data
of the proton electron scattering includes
a lot of absolutely different experiments. And second the dispersion
approach based on the theory of the analytic properties of the scattering
amplitude and other values involves a number of relation between the data
from other kinds of collision. For instance, the neutron data obtained from
scattering on nuclei are included in the evaluation.
\item The measurement of the $1s-2s$ transition frequency in the hydrogen
atom \cite{Udem} gives a result for the
ground state Lamb shift and hence for the proton radius. The
precise investigation of the hydrogen spectra involves
laser spectroscopy to measure the
transition frequencies and quantum mechanics and quantum
electrodynamics to calculate all contributions to the energy levels beside the
radius term.
\end{itemize}

These short explanations of the results demonstrate the universality of the
proton charge radius. This gives us the possibility to perform some
cross-checking of a number of different pieces of our knowledge on the atoms
and particles. There are also a number of other subfields where the
knowledge of the radius
is interesting, but they cannot give any precise result. An example of
those can be some models of the proton or neutron. As some important
applications the proton polarizability and the empiric determination of the neutron
magnetic radius can be mentioned.

The results are given in Table 1 and Fig.1 within the chronological order,
but we will consider them in detail starting with the Lamb shift
measurement. Next we will discuss the lattice calculations and then the
scattering data and their different fitting. As one of
the applications the hydrogen hyperfine structure and so called polarizability
contribution is considered in Appendix.

\section{The hydrogen Lamb shift for the proton radius}

The nuclear dependent correction to the Lamb shift in the hydrogen atom as it is well
known is of the form\footnote{We use a convention: $\hbar=c=1$,
$\alpha=e^2$. The nuclear charge $Z$ is useful to classify the QED
contributions and in the hydrogen atom indeed one has $Z=1$.
We always give expressions for $E$ as for the energy,
but the numerical values (in kHz) are given for the
corresponding frequency value $
f=E/h=E/2\pi$.}
\[\Delta E (nl) =
\frac{2}{3}\,\frac{(Z\alpha)^4}{n^3}\, m^3 \, R_p^2\;\delta_{l0}\;
.\]
If the energy were measured with high accuracy and all other contributions
to energy were known one could extract the proton charge radius from
hydrogen spectroscopic data. However, it is  only possible to measure some
differences of the state energies: either splittings or transition
frequencies.
They are dimensional values and to determine them one has to use some other
dimensional reference values. We give a review of all experimental results,
but before that we would like to attract the attention to a problem appeared
recently due to a significant progress in the optical measurements.

There are two kinds of the optical measurements from which one can extract the value
of the Lamb shift. First let us consider absolute optical measurements. They
have been done for the $1s-2s$, $2s-8s$ and $2s-8d$ transitions. One
problem
to utilize  their result is unknowledge of the Rydberg constant which mainly
determines the value of the transition frequency. The Lamb shift may be
found only by
using experimental results for two or more transition and solving a system of
equations where the constant is one of the variable to find. In this
way one gets the value of the Lamb shift from some special difference of
two frequencies where the Rydberg constant contribution is canceled:
\beq \label{DifFr}
\left(\frac{1}{2^2}-\frac{1}{8^2}\right)\cdot E(1s-2s)-
\left(\frac{1}{1^2}-\frac{1}{2^2}\right)\cdot E(2s-8s/d)\;.
\eeq

The other method of optical measurements (so called `ground state Lamb
shift measurements') is based on producing difference like in \eq{DifFr}
experimentally as a beat frequency. Measurements of $1s-2s/2s-4s$, $1s-2s/2s-4p$,
$1s-2s/2s-4d$, $1s-3s/2s-6s$ and
$1s-3s/2s-6d$ transitions have been done.

One can see that in the optical measurements it is not possible to measure the
Lamb shift of one separated level. Usually they can determine something
like
\[
\Delta E_L(1s) - C \Delta E_L(2s) + \dots\;,
\]
where the known constant $C$ depends on
the experiment and usually it lies between 4 and 5. The points stands for the Lamb
shift of higher $s$ levels and of levels with higher orbital momentum $l>0$.
All $s$ level energy depend on the proton radius.

A way of evaluating the data in which
a combination of Lamb shifts of different $s$-states is involved
has been considered in our works \cite{Karshenboim94,Karshenboim95a}. It is
possible to recalculate all $ns$ shifts to the ground state value by
using an auxiliary  difference
\[
\Delta(n)  = \Delta E_L(1s) - n^3 \Delta E_L(ns)\;,
\]
which is radius-independent. The details of this calculation
can be found in our paper \cite{Karshenboim97a}.

The higher $l$ levels are not radius-dependent on this level of
accuracy and they can be also calculated with good enough precision
\cite{Karshenboim94,Karshenboim95a}.

\subsection{Status of the Lamb shift theory}

We now  can compare the status of the ground state Lamb shift theory and the
$\Delta(n)$ theory. All recent results for the $1s$ (or $2s$) Lamb shift are
presented in Table 2.
In the Table we have included only results obtained directly for the hydrogen
atom.
However, it has to be mentioned that some works for hydrogen-like ions
with the nuclear charge $Z$
can be useful also for $Z=1$. Particularly, some important numerical results
were obtained in works \cite{Mohr} ($\alpha\,m$, all order of $(Z\alpha)$,
$Z=5,10...$) and \cite{Artemyev} ($m^2/M$, all order of $(Z\alpha)$,
$Z=1,2...$).

\begin{table}
\begin{center}
\begin{tabular}{||c|c||}
\hline\hline
&\\[-1ex]
Order                          & Reference  \\  [1ex]
\hline
\hline
&\\[-1ex]
$\alpha(Z\alpha)^6 m$ & Pachucki, 1993,
  \protect{\cite{Pachucki93}}           \\ [1ex]
$\alpha(Z\alpha)^7 m \log{Z\alpha}$ & Karshenboim, 1994,
  \protect{\cite{Karshenboim94}}\\ [1ex]
$\alpha^2(Z\alpha)^6 m\log^3{Z\alpha}$ &  Karshenboim, 1993,
  \protect{\cite{Karshenboim93}}\\ [1ex]
$\alpha^2(Z\alpha)^5 m$ & Pachucki, 1994,
  \protect{\cite{Pachucki94}}           \\ [1ex]
 & Eides, Shelyuto, Grotch, 1995,
  \protect{\cite{Eides95}}           \\ [1ex]
$(Z\alpha)^6 m^2/M \log{Z\alpha}$ & Doncheski, Grotch, Erickson, 1991,
  \protect{\cite{Doncheski}}   \\ [1ex]
                    & Khriplovich, Milshtein, Yelkhovsky, 1992,
  \protect{\cite{Khriplovich}}   \\ [1ex]
                    & Fell, Khriplovich, Milshtein, Yelkhovsky,1993,
  \protect{\cite{Fell}}   \\ [1ex]
                    &  Pachucki, Grotch, 1995,
  \protect{\cite{Pachucki95a}}   \\ [1ex]
                    $(Z\alpha)^6 m^2/M$ & Pachucki, Grotch, 1995,
  \protect{\cite{Pachucki95a}}   \\ [1ex]
                    & Yelkhovsky, 1996,
  \protect{\cite{Yelkhovsky96}}   \\ [1ex]
                    & Eides, Grotch, 1997,
  \protect{\cite{Eides97}}   \\ [1ex]
                    & Yelkhovsky, 1997,
  \protect{\cite{Yelkhovsky97}}   \\ [1ex]
$\alpha(Z\alpha)^5 m^2/M$ & Bhatt, Grotch, 1987,
  \protect{\cite{Bhatt}}   \\ [1ex]
 & Pachucki, 1995,
  \protect{\cite{Pachucki95b}}   \\ [1ex]
 & Eides, Grotch, 1995,
  \protect{\cite{Eides96}}    \\ [1ex]
\hline\hline
\end{tabular}
\vspace{5mm}
\caption{New results for the $1s$ and $2s$ Lamb shift \label{t2}
}
\end{center}
\end{table}

One can see from Table 2 that some corrections were investigated by a number of
authors and it has to be mentioned that the results were  not always
in agreement. For instance it was a long and
dramatic history of the calculation of the logarithmic corrections
$(Z\alpha)^6 m^2/M \log{Z\alpha}$ which after all were found to be equal to
zero \cite{Fell}. Two actual contradictions of the ground state Lamb shift are
presented in Table  3. For the interpretation of the numerical results of
Ref.
\cite{Artemyev}, which include also all higher order corrections, it is necessary
to remember that uncertainty from the higher order corrections can be
estimated as the
$(Z\alpha)^5m^2/M$ term multiplied by $(Z\alpha)^2\log{Z\alpha}$, or on level of
1 kHz.

\begin{table}
\begin{center}
\begin{tabular}{||c||c|c||}
\hline\hline
&&\\[-1ex]
Order &  Reference & Contribution  \\
\hline \hline
&&\\[-1ex]
$(Z\alpha)^6m^2/M$ & Pachucki, Grotch,  1995  & -7.4~kHz \\[1ex]
 &  Artemyev, Shabaev, Yerokhin,  1995 & -7.1(9)~kHz \\[1ex]
 & Yelkhovsky,  1996 & 2.8~kHz \\[1ex]
 & Eides, Grotch, 1997  & -7.4~kHz \\[1ex]
 &  Yelkhovsky, 1997 & -16.5~kHz \\[1ex]
\hline
&&\\[-1ex]
$\alpha(Z\alpha)^5m^2/M$ & Bhatt, Grotch, 1987  & -20.3~kHz \\[1ex]
 &  Pachucki,  1995 & -13.9~kHz \\[1ex]
 & Eides, Grotch, 1995  & -20.7~kHz \\[1ex]
\hline\hline
\end{tabular}
\vspace{5mm}
\caption{Theoretical discrepancies for the $1s$ Lamb shift
\label{t3}
}
\end{center}
\end{table}

Now we can consider the status of the specific difference $ \Delta(n) = \Delta
E_L(1s) - n^3 \Delta E_L(ns)$ and the Lamb shift of the $p$ states.
One can see from Table 4, that the theoretical status of $\Delta(n)$ and $p$
state energy levels is physically more clear than the status of the
ground state Lamb shift. Some details can
be found in Refs.
\cite{Karshenboim94,Karshenboim95a,Karshenboim96a,Karshenboim97a,Jentschura}.

\begin{table}
\begin{center}
\begin{tabular}{||c|c|c|c||}
\hline\hline
&&&\\[-1ex]
Order                          & $E_L(1s)$ & $\Delta(n)$ & $E_L(np)$ \\  [1ex]
\hline
\hline
&&&\\[-1ex]
$\alpha(Z\alpha)^6 m$ & coefficient $\approx 30$ & coefficient $\approx 1$
& coefficient $\approx$ -1         \\ [1ex]
$\alpha(Z\alpha)^8 m \log^3{Z\alpha}$ & unknown & 0   & 0       \\ [1ex]
\hline
&&&\\[-1ex]
$\alpha^2(Z\alpha)^6 m\log^3{Z\alpha}$ &  30 kHz & 0  & 0         \\ [1ex]
$\alpha^2(Z\alpha)^6 m\log^2{Z\alpha}$ & unknown &  known  & known  \\ [1ex]
\hline
&&&\\[-1ex]
$\alpha^3(Z\alpha)^4 m$ & unknown & 0 & known             \\ [1ex]
\hline
&&&\\[-1ex]
$(Z\alpha)^6 m^2/M$ & disagreement  & 0    & agreement        \\ [1ex]
$\alpha(Z\alpha)^5 m^2/M$ &  disagreement  & 0  & 0             \\ [1ex]
\hline
&&&\\[-1ex]
$(Z\alpha)^4m (m R_p)^2$ & radius-dependence  & 0   & 0            \\ [1ex]
\hline\hline
\end{tabular}
\vspace{5mm}
\caption{Comparison of status of $E_L(1s)$ and
$ \Delta(n)$
\label{t4}
}
\end{center}
\end{table}

The general expression of the difference $\Delta(n)$ is of the form \cite{Karshenboim97a}
\[
\Delta(n) =
\frac{\alpha(Z\alpha)^4}{\pi}\frac{m_r^3}{m^2}\times
\Bigg\{- \frac{4}{3}\log{\frac{k_0(1s)}{k_0(ns)}}
\left(1+Z\frac{m}{M}\right)^2 +
\]
\[
(Z\alpha)^2\times\Bigg[\left(4\big(\log{n}-
\psi(n+1)+\psi(2)\big)-
\frac{77(n^2-1)}{45 n^2}\right)
\log{\frac{1}{(Z\alpha)^2}}+
A^{VP}_{60}(n)+G^{SE}_{n}(Z\alpha)\Bigg]
\]
\[
-\frac{14}{3}\frac{Z\,m}{M} \left(\psi(n+1)-\psi(2)
-\log{n}+\frac{n-1}{2n}\right)\Bigg\}
+
\frac{\alpha^2 (Z\alpha)^6m}{\pi^2}
\log^2{\frac{1}{(Z\alpha)^2}}\,B_{62}
,
\]
\noindent where $\log{k_0(ns)}$ is the Bethe logarithm and $\psi(z) = (d/dz)\log\Gamma(z)$.

\noindent Here:
\begin{itemize}
\item $G^{SE}_n(Z\alpha)$ is the {\sl one\/}-loop {\sl self-energy\/}
correction of order
$\alpha(Z\alpha)^6 m$ and  higher;
\item $A^{VP}_{60}(n)$ is the $\alpha(Z\alpha)^6 m$-contribution of
the {\sl vacuum polarization\/}  \cite{Ivanov97};
\item $B_{62}$ is the {\sl leading\/} logarithmic {\sl two\/}-loop correction
coefficient \cite{Karshenboim94,Karshenboim95b,Karshenboim96a}.
\end{itemize}

The final results for the difference $\Delta(n)$ in the hydrogen and
deuterium atoms as well as for their isotopic shift were found in Ref.
\cite{Karshenboim97a} and they are summarized in Table 5.

\begin{table}
\begin{center}
\begin{tabular}{||c|c|c|c||}
\hline\hline
&&&\\[-1ex]
$n$ & $\Delta^{Hyd}(n)$ & $\Delta^{Deu}(n)$  &$\Delta^{Iso}(n)$\\[1ex]
 & [kHz] & [kHz] & [kHz] \\[1ex]
\hline
&&&\\[-1ex]
2 &  -187232(5.5)  & -187225(5.5) & 7.3 \\[1ex]
3 &  -235079(10) & -235073(10)& 5.9 \\[1ex]
4 &  -254428(12) & -254423(12)& 4.7 \\[1ex]
5 &  -264162(15) & -264158(15)& 4.0 \\[1ex]
6 &  -269747(15) & -269743(15) & 3.5\\[1ex]
7 &  -273246(16) & -273243(16) & 3.2\\[1ex]
8 &  -275583(17) & -275580(17) & 3.0\\[1ex]
9 &  -277221(18) & -277218(18) & 2.9\\[1ex]
10 & -278413(19) & -278410(19) & 2.7\\[1ex]
11 & -279308(19) & -279305(19) & 2.6\\[1ex]
12 & -279996(20) & -269993(20) & 2.5\\[1ex]
\hline\hline
\end{tabular}
\vspace{5mm}
\caption{Results of $\Delta(n)$ for $n=2\dots12$
\label{t5}}
\end{center}
\end{table}

By using the differences considered above one can easily recalculate the
Lamb shift of the $1s$ state to the Lamb shift of the $2s$. In this work we
prefer to discuss the Lamb splitting between the $2s_{1/2}$ and
$2p_{1/2}$ states.
Our motivation to do that is the possibility to compare the data
evaluated with $\Delta(n)$ (the optical measurement) to
the one
which obtained directly (the radiofrequency measurement).

Above we have compared the status of the $2s$ Lamb shift and the difference
$\Delta(n)$ and given the complete theoretical expression for the second \cite{Karshenboim97a}.
The theory of the $2s$ Lamb shift is much more complicated. We mainly agree
with a consideration in review \cite{Pachucki96}. However, it has to be
mentioned that compared to it,
our result is shifted by -3.6 kHz because the
$\alpha^2(Z\alpha)^6 m\log^3{Z\alpha}$-correction \cite{Karshenboim93}
has not been included there. We would  also like to explicitly present here
the sources of the theoretical uncertainty:
\begin{itemize}
\item[$\bullet$] unknown $\alpha(Z\alpha)^7 m$ and higher order corrections
are estimated to 1 kHz;
\item[$\bullet$] $\alpha^2(Z\alpha)^6 m \log^2{Z\alpha}$ and higher order
terms can give up to 2 kHz;
\item[$\bullet$] the $\alpha^3(Z\alpha)^5 m$ contributions previously
estimated here to 2
kHz need more understanding.
\end{itemize}

\subsection{Status of the Lamb shift experiment}

\begin{figure}[h]
\epsfxsize=10cm
\centerline{\epsfbox{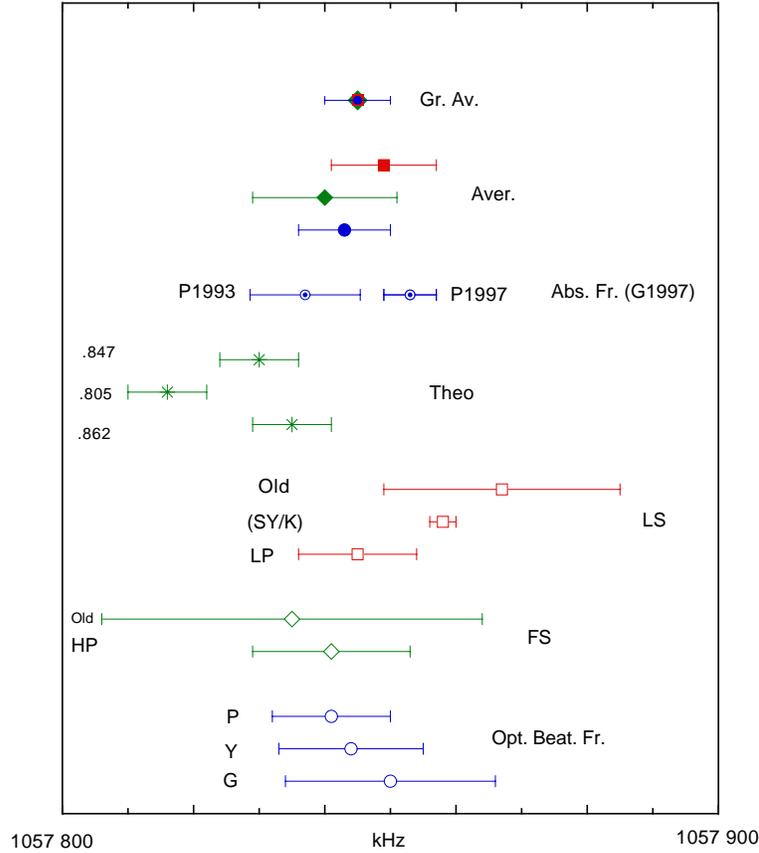}}
\caption{\label{f2} Different values of the Lamb splitting
$E(2s_{1/2})-E(2p_{1/2})$}
\end{figure}

All important result for the Lamb splitting ($2s_{1/2}-2p_{1/2}$) are
presented in Fig. 2. It is helpful first to describe all labels in the text and
next to discuss them. The references (and the transitions) of the experiments and
theory presented in the picture are:
\begin{itemize}
\item[{\sl Opt.Beat.Fr.}]{\bf Optical beat frequency measurements} were
realized with a high accuracy only recently:
\begin{itemize}
\item[{\bf G}] Measurement of $1s-2s/2s-4s$, Garching, 1994:
\cite{Weitz94}.
\item[{\bf  Y}] Measurement of $1s-2s/2s-4p$, Yale, 1993:
\cite{Berkeland}.
\item[{\bf  P}] Measurement of $1s-2s/2s-4p$, Paris, 1996:
\cite{Bourzeix96}.
\end{itemize}
\item[{\sl FS}]{\bf Fine structure measurements} deal
with the radiofrequency splitting ($2p_{3/2}-2s_{1/2}$).
\begin{itemize}
\item[{\bf HP}] The best and most recent result is of
{\sl E. W. Hagley and F. M. Pipkin, 1994} \cite{Hagley}.
\item[{\bf Old}] Older experiments were not very precise
and it is more convenient to consider them altogether. They include
results obtained by
\begin{itemize}
\item[$\circ$] {\sl B. L. Cosens and T. V. Vorburger, 1969} \cite{Cosens69};
\item[$\circ$] {\sl T. W. Shyn et al., 1971} \cite{Shyn71};
\item[$\circ$]
{\sl K. A. Safinya et al., 1980} \cite{Safinya} (result corrected
according to
{\sl E. W. Hagley and F. M. Pipkin, 1994} \cite{Hagley}).
\end{itemize}
\end{itemize}
\item[{\sl LS}]{\bf Lamb splitting measurements} have been performed directly
for the classical splitting $2s_{1/2}-2p_{1/2}$.
\begin{itemize}
\item[{\bf LP}] The result by {\sl S.~R.~Lundeen and F.~M.~Pipkin, 1981}
\cite{Lundeen81} is one of two relatively recent ones.
\item[{\bf SY/K}] The other experiment was performed by
Sokolov and Yakovlev and the result corrected by us.
The experiment properly by
{\sl Yu.~L.~Sokolov and V.~P.~Yakovlev, 1982} \cite{Sokolov82}
gives a result for the ratio of the $2p_{1/2}$ decay rate and the splitting.
The correction to decay rate was found in Ref.
\cite{Karshenboim94,Karshenboim95c} by
{\sl S. G. Karshenboim, 1994}. See also
{\sl Comments to  {\it On the Accuracy of Lamb Shift Measurements in Hydrogen},
(Physica Scripta,  55 (1997) 33--40)
by V. G. Pal'chikov, Yu. L. Sokolov, and V. P. Yakovlev} \cite{Karshenboim97b}.
\item[{\bf Old}] Older experiments are summarized
in one point. The results taken into account include
values obtained by:
\begin{itemize}
\item[$\circ$] {\sl D.~A.~Andrews and G.~Newton, 1976} \cite{Andrews};
\item[$\circ$] {\sl R.~T.~Robiscoe and T. W. Shyn, 1970} \cite{Robiscoe70}.
\end{itemize}
\end{itemize}
\item[{\sl Theo}] Some {\bf theoretical} values are presented too.
Different results appear due to the different
values of the proton radius discussed above. They are labeled with the proper
radius
values (see Table \ref{t1}):
\begin{itemize}
\item[{\sl .862}] -- Mainz scattering result \cite{Simon};
\item[{\sl .805}] -- Stanford scattering result \cite{Hand};
\item[{\sl .847}] -- Mainz dispersion fitting result \cite{Mergell}.
\end{itemize}
\item[{\sl Abs.Fr.(G1997)}] The {\bf comparison of absolute optical
frequencies} is presented separately from other experimental points
because in contrast to them it contains data from {\sl two\/} independent
experiments. The other reason is that this gives now the results with
the highest accuracy (except the Sokolov--Yakovlev result which
is discussed below).
The $1s-2s$ measurement \cite{Udem} done
in Garching (1997) cannot be used alone. The other absolute measured value used to
find the Lamb shift is a Paris result for the $2s-8s/8d$ frequency. We give
two points due to two different Paris experiment:
\begin{itemize}
\item[{\sl P1993}] Earlier result: Paris--1993 \cite{Nez93},
\item[{\sl P1997}] Recent result: Paris--1997 \cite{Beauvoir97}.
\end{itemize}
\item[{\sl Aver}]  It is also convenient to compare results of
{\sl Abs.Fr.(G1997)} with some {\bf average values}. We give average values
for three kind of experiments. The labels are easily to recognize.  We
would like to comment them:
\begin{itemize}
\item[{\sl OBF}] {\sl The
evaluation is performed with by $\Delta(n)$ \cite{Karshenboim97a} .}
\item[{\sl FS}] {\sl The result is obtained without using $\Delta(n)$,
but with applying the theoretical results for the fine structure
interval $2p_{1/2}-2p_{3/2}$ \cite{Jentschura}.}
\item[{\sl LS}] {\sl The result is obtained without using either
$\Delta(n)$ or the fine structure $2p_{1/2}-2p_{3/2}$.}
The {\sl SY/K}-value is not taken into account here.
\end{itemize}
\item[{\sl Gr.Av.}] The {\bf grand average value} is found over the
three average values mentioned above.
\end{itemize}

Before a detailed discussion we would like to attract the attention to
{\bf an important point}. As it is mentioned the difference $\Delta(n)$
has been involved into evaluation of optical beat frequency ({\sl Opt.Beat.Fr})
and
the absolute frequency ({\sl Abs.Fr.}) measurements the data. The fine
structure ({\sl FS}) and the Lamb splitting ({\sl LS}) measurements data are needed
no results of $\Delta(n)$. One can see in Fig. 2 that these two group
of the results ($\Delta(n)$-dependent and $\Delta(n)$-independent) are
in fair agreement.

Now we are listing some {\sl problems for the hydrogen measurements
adjustment\/} needed to be solved for extracting the proton radius from
the Lamb shift data.
\begin{itemize}
\item The first problem to solve is to fix the experiments
to be included and the uncertainties to be used for them.
\begin{itemize}
\item[$\circ$] Sokolov and Yakovlev experiment \cite{Sokolov82}
and its uncertainty is an open question because some criticism of Hinds
\cite{Hinds};
\item[$\circ$] Older experiments altogether can give an uncertainty of about
10 kHz. Not all of them are taken into account in Fig. 2. We would like here
to mention especially some Lamb splitting experiments and one for the fine
structure.
\begin{itemize}
\item[$-$] The value of the Lamb splitting \cite{Cosens69} is included and
the uncertainty ($\delta =
42\, kHz$) is taken by us according to the original work
in contrast to a consideration by {\sl B. N. Taylor et al., 1969}
\cite{Taylor69} (the uncertainty there is $\delta = 64\, kHz$).
\item[$-$] The result of the Lamb splitting by {\sl S. Triebwasser et al., 1953}
\cite{Triebwasser} is excluded because
 the work includes results for two lines. Being corrected
(see \cite{Taylor69} for details) they are
inconsistent;
$\delta = 64\,
kHz$ (according {\sl B. N. Taylor et al.} \cite{Taylor69}).
\item[$-$] The experiment by {\sl Yu.~L.~Sokolov, 1973} \cite{Sokolov73}
for the Lamb splitting is not included  being in straight disagreement with all
other measurements ($\delta = 40\, kHz$).
\item[$-$] The result for the fine structure by {\sl S. L. Kaufman et al., 1971}
\cite{Kaufman} is excluded also being in straight disagreement with
all other measurements ($\delta = 40\, kHz$).
\end{itemize}
\end{itemize}
\item One can see from the picture that the two absolute measurement of the $2s-8d/8s$
transition from Paris are in a disagreement:
the earlier result of 1993 \cite{Nez93}  is
\[
f(2s-8d_{5/2})=770\, 649\, 561 5{\bf 67}(10)\; kHz
\]
and the recent result of 1997: \cite{Beauvoir97} is
\[
f(2s-8d_{5/2})=770\, 649\, 561\, 5{\bf 85}(5)\; kHz.
\]
Actually these are two different experiment
where different standards have been used and the results  are expected to be
partly
independent. In another work \cite{Nez95} the older experiment
result is presented as
\[
f(2s-8d_{5/2})=770\, 649\, 561\, 5{\bf 71}(12) \; kHz.
\]
It is not clear if we have to use one of them (which one ?) or to
find some average value.
\item It has to be also mentioned that actually there are a number
of older less precise measurements
from Paris  which are quite correlated.
\item Another question that appears for such high accuracy as 3 kHz
for the $2s$ Lamb shift is the fine structure constant to be used.
The different choice  of it can lead to shift of the $E(2s_{1/2}) -
E(2p_{1/2})$ splitting value up to 1
kHz for the results obtained by
the {\sl optical beat frequency} and {\sl absolute frequency}
measurements.  For the
$E(2s_{1/2}) - E(2p_{1/2})$ value from the {\sl fine structure}
measurements the shift
is up to -2 kHz, where `--' indicates that the shift is in the opposite
direction to the one for the results extracted from the
{\sl Opt.Beat.Fr.} and {\sl Abs.Fr.} experiments.
The results for the direct measurements of {\sl Lamb splitting}
cannot be significantly shifted. That is because all changes due to
correcting of the fine structure constant
value come via the $(Z \alpha)^4\,m$ term
(the relativistic correction to the Schr\"odinger energy which is
also responsible for the hydrogen fine structure). This term is about an order
of magnitude {\bf larger} than the Lamb shift properly.
Due to these different signs the {\sl grand average} value is less
sensitive to the choice of $\alpha$, but according to this choice the set of
results may be more or less inconsistent.
\item One can combine the data and extract the value of the fine structure
constant. The uncertainty is of the level of $3\cdot 10^{-7}$ and it seems
that it cannot be reduced in the nearest future.
\end{itemize}

Some new problem can appear when taking into account {\sl the deuterium
results}. It is important to remember:
\begin{itemize}
\item Isotopic shift of the $1s-2s$ transition frequency
has been measured in Garching within {\bf 0.15} kHz
uncertainty \cite{Huber}.
\item Isotopic shift of $\Delta(n)$ with uncertainty less than {\bf 0.05} kHz
has been obtained by {\sl S. G. Karshenboim, 1997\/} \cite{Karshenboim97a} (see Table
\ref{t5}).
\item That gives the isotopic shift
  \begin{itemize}
\item [$\circ$] of the $1S$ Lamb shift with an uncertainty of {\bf  0.17} kHz;
\item [$\circ$] of the $2S$ Lamb shift with an uncertainty of {\bf  0.02} kHz.
   \end{itemize}
\item The list of the deuterium results similar to the case of the hydrogen
has to include a number of old and recent ones. They are shortly discussed
below\footnote{The items marked by `$-$' should rather be excluded
from adjustment in contrast to ones with `$\circ$' which should be taken into
account.}.
     \begin{itemize}
\item[$\circ$] The measurement of the Lamb splitting by
{\sl B. L. Cosens, 1968} \cite{Cosens68} has been
correlated with the hydrogen one ($\delta=33\;kHz$).
\item[$-$] The experiment by  {\sl S. Triebwasser et al., 1953}
\cite{Triebwasser} for
the Lamb splitting also has contained correlations
with the hydrogen result and
leaded to the same problem as for hydrogen: the
results for two lines corrected according to {\sl B. N. Taylor et
al., 1969} \cite{Taylor69} are inconsistent ($\delta=35\;kHz$).
\item[$-$] The work by
{\sl E. S. Dayhoff et al., 1953} \cite{Dayhoff} devoted to the
fine structure have contained a disagreement between the results for two lines
($\delta=53\;kHz$).
\item[$\circ$]  The optical beat frequency measurement of $1s-2s/2s-4s$
transitions in the deuterium atom (Garching, 1994,
\cite{Weitz94}) has been done with the same experimental setup
as for the hydrogen ($\delta=28\;kHz$).
\item[$\circ$] The best absolute measurement has been performed recently for the
$2s-8s/8d$ transitions in Ref.
\cite{Beauvoir97} (Paris, 1997) and it has been correlated with the hydrogen
measurement; the uncertainty is the same as for the hydrogen ($\delta=5\;kHz$).
\end{itemize}
\item One can see that {\bf all} deuterium results are more or less correlated
with the hydrogen results obtained by the same teams.
\item It seems that the Sokolov and Yakovlev experiment is the only direct way
where it could be possible to obtain uncertainty below 1 kHz for the 2s Lamb
shift. Due to a perfect reproductibility, the unknown systematic error can be
expected to be the same for hydrogen and deuterium.
\end{itemize}

\section{Lattice calculations}

One of the other methods that gives a value for the proton radius (see Table \ref{t1}
and Fig. 1) is the approach based on the lattice calculation within the chiral
limit in which the pion mass is equal to zero. We expect that the result of Ref.
\cite{Leinweber93} included in the Table 1 and Fig. 1 is not quite correct. Our
opinion is based on the following critical remarks:

\begin{itemize}
\item Small lattice ($24 \times 12\times 12\times 24$) has been used in the
calculation. Some progress is possible: in a recent evaluation of the neutron form
factor in Ref. \cite{Wilcox} a lattice $20^3\times 32$ was used. In case of
the small lattice only a not too large number of discrete momenta are available.
The proton has to occupy a relatively large portion of the volume of the whole space.
\item That is not an {\sl ab initio} calculation.
It includes a number of free auxiliary parameters found by a barions masses
fitting.
\item The chiral limit ($m_\pi$ = 0) also leads to some limitation
for the available values of the low momentum.
\item The finite space implies neither $q^2 = 0$ nor $q^2 \to 0$, because only some discrete
value of $q$ are allowed. The radius cannot be found directly, but only
by some fitting.
\item As far as the form factor $G_E(q^2)$ could be found only at not too small
$q^2$ (e.g. 0.16 {\rm Gev}$^2\simeq 4\,{\rm fm}^{-2}$), it has been fitted with monopole or dipole fits.
They have obtained results for only a few values of $q^2$ and used a
normalization at $q^2=0$.
\item The result looks very sensitive to chiral perturbating: the value
of the proton radius was shifted
from {\bf 0.64(8)}~{\rm fm} \cite{Draper} to
{\bf 0.88(3)}~{\rm fm} \cite{Leinweber93}. It may seem
quite strange that using the chiral correction a smaller uncertainty has been
obtained.
%
\item {\sl But a more important question is the status of their estimation
of the uncertainty}.
 \begin{itemize}
 \item [$\circ$]
Due to the simple Yukawa picture for the proton with the charged pion clouds it is
not quite clear how the proton charge radius can be finite in the massless pion limit.
The virtual neutral pions also lead to a divergency which is similar to
the well known
divergency in the electric form factor of the electron in the usual quantum
electrodynamics.
 \item [$\circ$] The proton radius is also expected to be
infinite in the pure chiral limit according to the correcting fit of Ref.
\cite{Leinweber93}.
 \item [$\circ$] But in the
previous work \cite{Draper} the same authors just used only
the chiral limit and obtained a finite result !
 \item [$\circ$] If they are not able to recognize an actual divergency
how one can believe their result and especially the uncertainty ?
 \item [$\circ$]
We expect that was possible maybe because
they used only a few of the values of not too low momenta $q^2$ and the dipole or
monopole fits. The divergency were actually cut off with the finite
volume of the space. It also has to be mentioned that the
normalization  $G(0)=1$ was also applied by them,
but it has no clear physical meaning because in finite space $q^2\neq 0$.
 \item [$\circ$] The
results are actually sensitive to the fitting procedure, which can allow to
obtain a finite value instead of infinity.
 \end{itemize}
\end{itemize}

\section{Scattering data}

We now can start to discuss the other traditional way to investigate the
proton radius based on the unpolarized elastic electron-proton scattering.
The cross-section described by the well known Rosenbluth formula is of the
most simple form in terms of the Sachs form factors:
\beq  \label{Rose}
\frac{d\sigma(E_0,\theta)}{d\Omega}=
\left[\frac{d\sigma(E_0,\theta)}{d\Omega}\right]_0\times
\left[A(q^2)\,G_E^2(q^2)
-\frac{q^2}{4M^2}\,B(q^2,\theta)\,G_M^2(q^2)\right]
\;,\eeq
where
\[
\left[\frac{d\sigma(E_0,\theta)}{d\Omega}\right]_0=
\frac{\alpha^2}{4E^2_0}
\frac{\cos^2\frac{\theta}{2}}{\sin^4\frac{\theta}{2}}
\frac{1}{1+\frac{2E_0}{M}\sin^2\frac{\theta}{2}}
\]
is the cross-section for scattering with a spinless point-like particle
(instead of the proton) and $q^2=q_0^2-{\bf q}^2<0$.

The coefficients in \eq{Rose} are known
\[
A(q^2)=\frac{1}{1-q^2/4M^2}
\]
and
\[
B(q^2,\theta)=\left(\frac{1}{1-q^2/4M^2}+
2\,\tan^2{\frac{\theta}{2}}\right)\;
.\]
In the limit of the small momentum
$q^2$ they are equal to
\[
A(0)=1\,
,\]
\[
B(0,\theta)=1+2\,\tan^2{\frac{\theta}{2}}\,
.\]

\bigskip

The proton (electric) squared radius in terms of the form factor is defined as
\beq  \label{RadFF}
R_p^2 =
\frac{1}{6\, G_E(0)}
\frac{\partial G_E(q^2)}{\partial\,q^2}\Bigg\vert_{q^2=0}\;
.\eeq

The Sachs form factors $G_E$ and $G_M$ are not only useful
in the particle physics. The others
for particle with spin 1/2 are the Dirac ($F_1$) and the Pauli
($F_2$) form factors. In their terms the vertex of the emission of a
virtual photon by a real proton is of the form
\[
\Gamma_\mu=F_1(q^2)\gamma_\mu+F_2(q^2)\left(-\sigma_{\mu\nu}
\frac{q^\nu}{2M}\right)\;.
\]
There are two relations between the different form factors:
\[
F_1(q^2)=\frac{G_E(q^2)-(q^2/4M^2) G_M(q^2)}{1-q^2/4M^2}
\]
and
\[
F_2(q^2)=-\frac{G_E(q^2)-G_M(q^2)}{1-q^2/4M^2}
\;.\]

\begin{figure}[h]
\epsfxsize=16cm
\centerline{\epsfbox{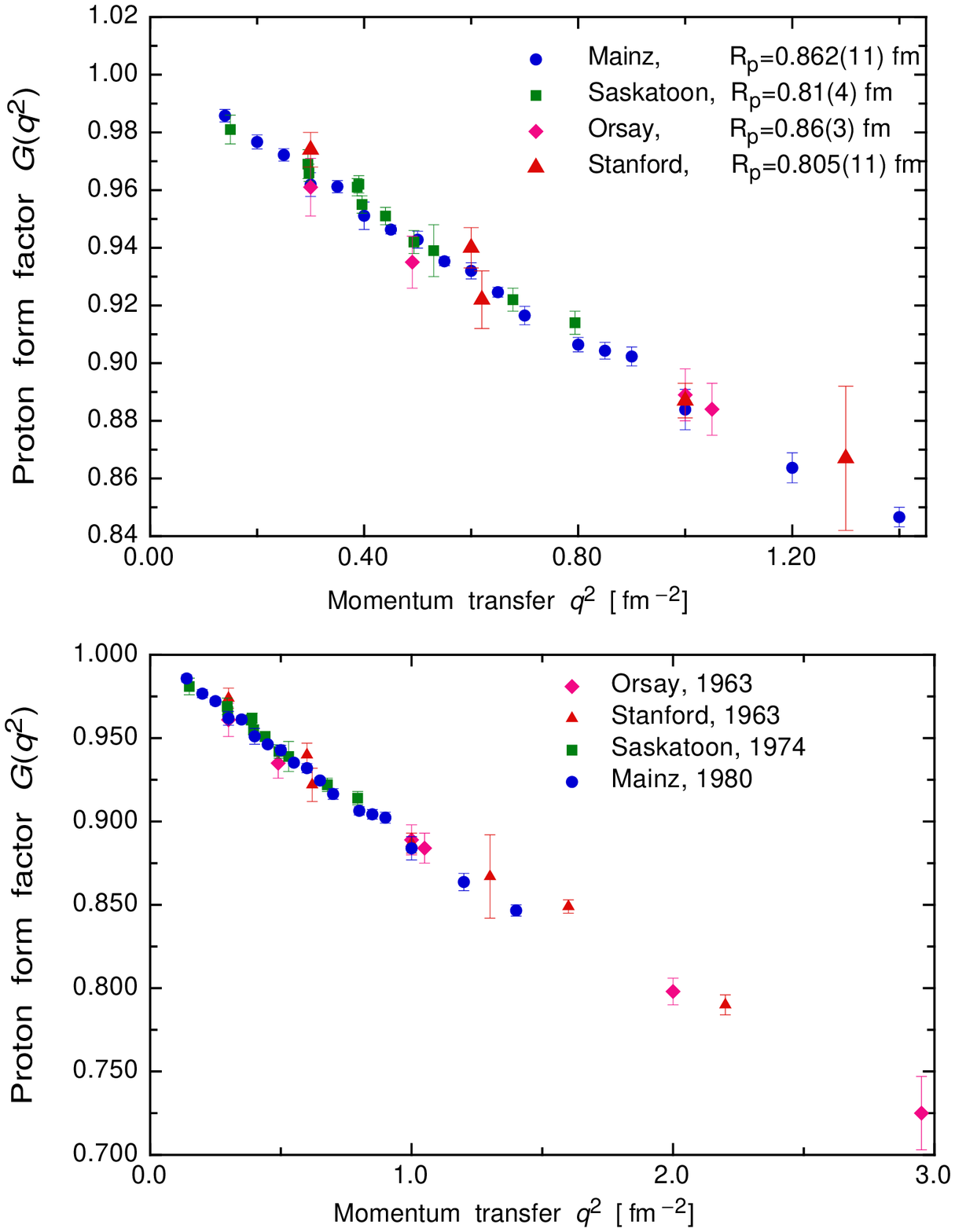}}
\caption{\label{f3} Experimental data of the elastic electron-proton
scattering are given for the whole
low energy range ($a$), and they are only for the Mainz-1980
fitting momentum range ($b$)}
\end{figure}

By measuring the scattering cross-section and applying the Rosenbluth formula
one can obtain the values of the form factors. A
number of problems appear when {\bf fitting the scattering data}.
     \begin{itemize}
\item{\sl Dipole approximation}.\\
The so called dipole approximation is of the form
\[
G_E(q^2)=\frac{G_M(q^2)}{1+\kappa}
= \left[\frac{\Lambda^2}{\Lambda^2-q^2}\right]^2
.\]
This is a very good approximation for any preliminary estimation, but it has no
theoretical status. It was used in some old fitting.
\item{\sl Extrapolation.}\\
In the Mainz experiment \cite{Simon} they used a three term extrapolation
\[
G_E(q^2)=\frac{G_M(q^2)}{1+\kappa}=a_0+a_1\,q^2+a_2\,q^4\,.
\]
The same was done in some older works too (see e. g. \cite{Murphy}). From
the charge normalization one could expect $a_0=1$. But actually the problem
is not so
simple and we discuss that below after consideraing the problem due to
the magnetic form factor.
\item{\sl Magnetic radius}.\\
In the Mainz-1980 evaluation of the data as well as in a lot of previous works
it a relation
\[
G_E(q^2)=\frac{G_M(q^2)}{1+\kappa}
\]
was used for fitting.
That definitely works for $q^2=0$
\beq \label{norma}
G_E(0)=\frac{G_M(0)}{1+\kappa}=1.
\eeq
and definitely does not work in general because, e. g.
\[
G_E(4M^2)=G_M(4M^2).
\]
Hence, this is only an approximation and it is not quite clear what
systematic errors come from the difference between $G_E(q^2)$ and
$G_M(q^2)/(1+\kappa)$. In the Mainz experimental work \cite{Simon}
they used for the evaluation only data for
which the magnetic effects are less than 10\%. That means that the uncertainty
for $R_M$ cannot be  smaller than 10 times $\delta R_E$, if we have
nothing else to estimate the difference $R_E-R_M$. In the Stanford work they used
also some higher momentum measurements. We give all accurate experimental points
for the electric form factor in two pictures. In Fig. 3b results are included for
range of momentum used in the Stanford fitting (up to ${\bf q}^2=3\,{\rm fm}^{-2}$), but
in Fig. 3a we keep only the same momenta as in the Mainz experimental fitting for
which magnetic effects can be neglected (${\bf q}^2\leq 1.4\,{\rm fm}^{-2}$). {\bf
Experimental status of the problem} can be illustrated with a result by {\sl H.
Treissen and W. Sch\"utz, 1974}
\cite{Treissen} done for the  not too high momenta $q^2$
\beq \label{Gratio}
G_E(q^2)=1.01(3)\cdot \frac{G_M(q^2)}{1+\kappa}.
\eeq
In case of a low momentum transfer one can write
\beq \label{EAs}
G_E(q^2)\approx 1 + \frac{R_E^2q^2}{6}
\eeq
and
\beq   \label{MAs}
G_M(q^2)\approx \big(1+\kappa\big)\,\left(1 + \frac{R_M^2q^2}{6}\right),
\eeq
and for all low energy of the Mainz data that they used for the fitting
it was
\[
1.5\cdot 10^{-2} < \Big(G_E-1\Big) \approx
\frac{R_M^2q^2}{6} < 15\cdot 10^{-2}\,.
\]
That means that for so low momenta $q^2$ the result of \eq{Gratio}
comes almost from the normalization \eq{norma} and actually cannot
give any information on the radii.

For high $q^2$ there are no direct
connection  between the radii and form factors, but for the low momenta
as we can see the
electric radius term is of the order $10^{-2}$.
Thus there is no significantly accurate result for
critical comparison of $R_E$ and $R_M$.
\item{\sl Normalization of cross section}.\\
Measuring cross section one has to take care of a proper
normalization factor. Due to
the experimental normalization the result of extrapolation
\beq \label{fit}
G_E(q^2)=a_0+a_1\,q^2+a_2\,q^4.
\eeq
to $q^2\to 0$
has {\bf not} to give $a_0$ equal to 1 exactly. That is not possible due to the
measurement nature of the normalization. But of course $a_0$ has to be in
agreement with 1 {\bf within the experimental uncertainty}.
And an more point:
the uncertainty mentioned has to be interpreted as the systematic error and
thus the value of $a_0$ can be different for different experiments. E. g. in the Mainz-1980 work
they used different values to fit the data of three experiments:
\begin{itemize}
\item $a_0(Mainz)=1.0014$ \cite{Simon};
\item $a_0(Orsay)=1.020$ \cite{Lehmann};
\item $a_0(Saskatoon)=1.008$ \cite{Murphy}.
\end{itemize}
The physical meaning of the constant $a_0$ is: {\sl the function $G(q^2)$
determined straightforwardly from the Rosenbluth formula
is\/} {\bf not} {\sl the true form factor}. The true one is $G(q^2)/a_0$.
The true value of radius has to be defined as
\[
R_p^2=\frac{a_1}{6\,a_0}.
\]
From the Mainz experiment article \cite{Simon}
it is not quite clear if they used the definition above
or
\[
R_p^2=\frac{a_1}{6}.
\]
The last equation is wrong. That can be important for incorporation of
the Orsay and Saskatoon data.\\
In the dispersion fitting work \cite{Mergell} they have not used different
normalizations for different experiments, but only
\[
G_E(0)=1.
\]
As a result they have underestimated the
uncertainty because all systematic errors are divided there by
a large statistical factor due to the number of the experimental points.
\end{itemize}

We would like to discuss here shortly a well known contradiction between the
Stanford \cite{Hand} and Mainz \cite{Simon} empirical fitting results for the
radius. The Mainz result was obtained by evaluating Mainz data properly
\cite{Simon},
Saskatoon data \cite{Murphy} and Orsay data \cite{Lehmann}. The Stanford radius
was found by
treating the Stanford \cite{Hand}, Orsay \cite{Lehmann} and some less precise
data. The last are not presented in Figs. 3. One can easily see from the
Fig. 3b, that the data used in the Mainz work are much anymore representative
and significantly
more precise. Some high momentum points in Fig. 3a cannot really change the
situation because they are not too accurate. We will not consider more the Stanford data here.

An other important point is due to the {\bf QED corrections}.
In order to determine the true proton form factor from the Rosenbluth formula
one has to take into account a number of such corrections:
\begin{itemize}
\item The main radiative corrections are due to the electron form factor and
the bremsstralung, and the electronic vacuum polarization. The next
important term is the two photon exchange. The complete result in the one loop
approximation was found in Ref. \cite{Mo}. However, in earlier experimental works
some incomplete results (see Refs. \cite{Meisner,Tsai}) were used. E. g. the QED
corrections according to Ref. \cite{Tsai} were used in papers \cite{Hand,Lehmann}, but in
the evaluation in work \cite{Murphy} the results of Ref. \cite{Meisner} were used.
Before including the data mentioned above to any compilation it is necessary
to correct them according to Ref. \cite{Mo}.
It is not clear if that has been done
or not in the both Mainz compilations.
\item Two-loop radiative corrections of the order $\alpha^2$ include
large squared logarithms $\log{(q/m_e)}$.
In Ref.  \cite{Mo} which is now used as the most
complete QED result the authors claimed that part of the two-loop corrections may
be taken into account using an exponential of the one-loop correcting factor
$\delta_1$
\[
1+\delta_1 \to exp\big(\delta_1 \big)
\,.\]
That may allow to include some leading logarithmic terms.
In earlier experimental works they were not
taken into consideration and it is not clear if they
have been actually included in the Mainz experiment evaluation \cite{Simon}.
In the Mainz-1980 work there is no prescription
how the QED corrections have been treated.
The two-loop correction may contribute on
the 1\% level of the
$\big(G-1\big)$ value for a few lowest momenta
$q^2$ of the Mainz experimental data. The result for the two-loop vacuum polarization
is well known and the two-loop form factor of the electron was found in Ref.
\cite{Barbieri}.
\item The muonic and hadronic vacuum polarization effects can be also
important. The muonic and hadronic vacuum polarization leads to a
correction lying in a range from
1\% to 0.5\% of the $G-1$ value for all experimental points in Ref.
\cite{Simon}. It has to be
compared with about the 1\% precision in the radius determination.
\item In Mainz-1980 there are no explicit equations for the Rosenbluth
scattering. In earlier works often the massless electron approximation $m_e=0$
was used. The correction is of the order $m_e^2/E_0^2$ and it has to be
$10^{-4}-10^{-5}$. It is not clear if older data are corrected and if in
Mainz-1980 data this approximation has been used or not. Being
$E_0$-dependent it
cannot be incorporated in the fit.
\end{itemize}

Thus our {\bf QED summary} contains two statement
\begin{itemize}
\item It is not clear if the older data from Refs. \cite{Lehmann,Murphy}
are included in the final Mainz compilations being corrected
according to Ref. \cite{Mo} or not.
\item All QED corrections which can be important on the 1\% level precision of
the charge proton radius determination have been known and they have to be
taken into account when fitting the data.
\end{itemize}

Now we would like to discuss some results from the {dispersion fitting
approach}. It allows to incorporate data with both high and low momentum transfer.
The results for electric and the magnetic proton radii as well as for the
magnetic neutron radius have uncertainties on the level of 1\% or 0.009 {\rm fm}
\cite{Mergell}. That seems strange because:
\begin{itemize}
\item The low energy fitting and high energy one can be done absolutely
separately.
\item It is clear that we have some additional information only
for the proton electric radius which comes from the low energy scattering
when the
magnetic form factor effects can be neglected. How can the uncertainties
be the same ?
\item The low energy scattering is expected according to
the empirical fit of Ref.
\cite{Simon} to  lead to {\bf 0.862(12)} {\rm fm}, and one can obtain the
high energy part
from re-evaluation back from the average value {\bf 0.847(9)} {\rm fm} \cite{Mergell}. It is close to {\bf
0.832(12)} {\rm fm} and that is it straight disagreement with the low energy result \cite{Simon}.
\end{itemize}

Concerning the second remark on the different uncertainties expected by us, we would
like to say that the fitting has been done in Ref. \cite{Mergell} for the
isoscalar
\[
F^S_i(q^2)=\frac{F^p_i(q^2)+F^n_i(q^2)}{2}
\]
and isovector
\[
F^V_i(q^2)=\frac{F^p_i(q^2)-F^n_i(q^2)}{2}\,
\]
form factors, combined from the Dirac ($i=1$) and Pauli ($i=2$) form factors of
the proton and neutron. We expect that the authors have obtained rather
results for the fitting parameters but they are not ready to estimate properly the
uncertainty for their specific function like the radii.

Due to the different data from the {\bf low and high energy scattering}
we first discuss the different physics of the correction.
\begin{itemize}
\item The {\sl two-loop QED correction\/} are of different magnitude for the
low and high momentum.
 \begin{itemize}
\item[$\circ$] The  {\sl low energy\/}  scattering is sensitive to them.
\item[$\circ$] The  {\sl high energy\/}  one is not.
 \end{itemize}
\item  The {\sl muonic and hadronic vacuum polarization\/}  are of the same relative
$1\%$ level for both of them.
 \begin{itemize}
\item[$\circ$] The  {\sl low energy\/}  data are sensitive.
\item[$\circ$] The  {\sl high energy\/}  ones are sensitive as well.
 \end{itemize}
\item  The {\sl analytic properties\/} of the form factors can be of use.
 \begin{itemize}
\item[$\circ$] For the  {\sl low energy\/} data evaluation they are not needed.
It is a pure empirical fitting
\item[$\circ$] The  {\sl high energy data\/}
treatment in Ref. \cite{Mergell} has been actually based on their using.
 \end{itemize}
\item  The {\sl normalization of the experimental data\/}  is the
of one important points as we show below.
 \begin{itemize}
\item[$\circ$] The  {\sl low energy\/}  data are sensitive to that up to
the level of $10^{-4}$.
 \begin{itemize}
\item In the Mainz-1980 work \cite{Simon} it is included as a fixed parameter
with the different values for different experiments.
\item In the Mainz-1996 paper \cite{Mergell} the normalization of the
electric form
factor is fixed absolutely to be equal to 1 at $q^2=0$ . In this way one has
to take care of the systematic
errors due to correlation of the data from the same experiment.
 \end{itemize}
\item[$\circ$] In case of the  {\sl high energy\/}  scattering, the data are sensitive to
the normalization only on level up to $10^{-2}$.
 \end{itemize}
\item  The {\sl magnetic form factor effects\/}  also have to be mentioned.
 \begin{itemize}
\item[$\circ$] The  {\sl low energy\/}  scattering is a mainly pure electric form factor
scattering and magnetic effect are small.
\item[$\circ$] The  {\sl high energy\/}  data include such effects significantly.
 \end{itemize}
\end{itemize}

The short {\bf summary} of the comparison `{\bf low energy against the high
one}' can be described with following listing:
 \begin{itemize}
\item The {\sl low energy\/}  data are
 \begin{itemize}
\item[$-$] sensitive to everything;
\item[$+$] needed no additional theory (e. g. like the dispersion approach);
\item[$\pm$] only for $G_E$.
 \end{itemize}
\item The {\sl high energy\/}  data are
 \begin{itemize}
\item[$+$] not so sensitive;
\item[$-$] needed to be treated with using significantly a knowledge of the
analytic properties;
\item[$\pm$] both for the electric and magnetic form factors.
 \end{itemize}
 \end{itemize}
The items labeled by `$+$' are advantages and the ones with `$-$' are
disadvantages.

Now after the comparison of the low energy and high energy data
we would like to discuss some {\bf interpretation of
the double Mainz difference}. We have reported on some possible disagreement
between the low energy and high energy data, but actually there are two options:
 \begin{itemize}
\item The disagreement can be actual an one and it has been discussed some possible
corrections which are different for the high and low energy data.
\item In the new Mainz work \cite{Mergell} they could obtain the
different result from the one in Ref. \cite{Simon} for
the low energy part of the data, because they could use another fitting.
Some different
details of their approaches have been also discussed above.
 \end{itemize}
Unfortunately in Ref. \cite{Mergell} the authors have given no discussions
which part of the data is mainly statistically responsible for the final
result of the proton radius. If we suppose that the difference
appeared from the low energy data, the
main source of them is the experimental results of Ref. \cite{Simon}.
Maybe some clarifying remarks have been just done by Wong (ironically his
work \cite{Wong} was appeared in 1994 before the
publication of the paper \cite{Mergell} two years later). In the quoted work,
only the Mainz data
have been fitted according \eq{fit}, but in three different ways.
We present all of them in Fig. 4.
 \begin{itemize}
\item The normalization constant is fixed to $a_0=1.0014$. The result is
$R_p={\bf 0.863(9)} $ {\rm fm} close to $R_p={\bf 0.862(12)} $ {\rm fm} from
Mainz experimental paper \cite{Simon}.
The procedure is also close to the one of Ref. \cite{Simon}.
The difference may come from including the data of the two other less
precise experiments or from a different treatment of the systematic errors.
\item The normalization constant is fixed to $a_0=1$. The result is
$R_p={\bf 0.849(9)} $ fm close to $R_p={\bf 0.847(9)} $ {\rm fm} from
the dispersion fitting work \cite{Mergell}. The same
normalization was used there.
\item The normalization parameter $a_0$ has been
treated as a free parameter to be
found from the fit. The results are $a_0=1.0028(22)$ and $R_p={\bf
0.877(24)}$ {\rm fm}.
 \end{itemize}
\begin{figure}[h]
\epsfxsize=16cm
\centerline{\epsfbox{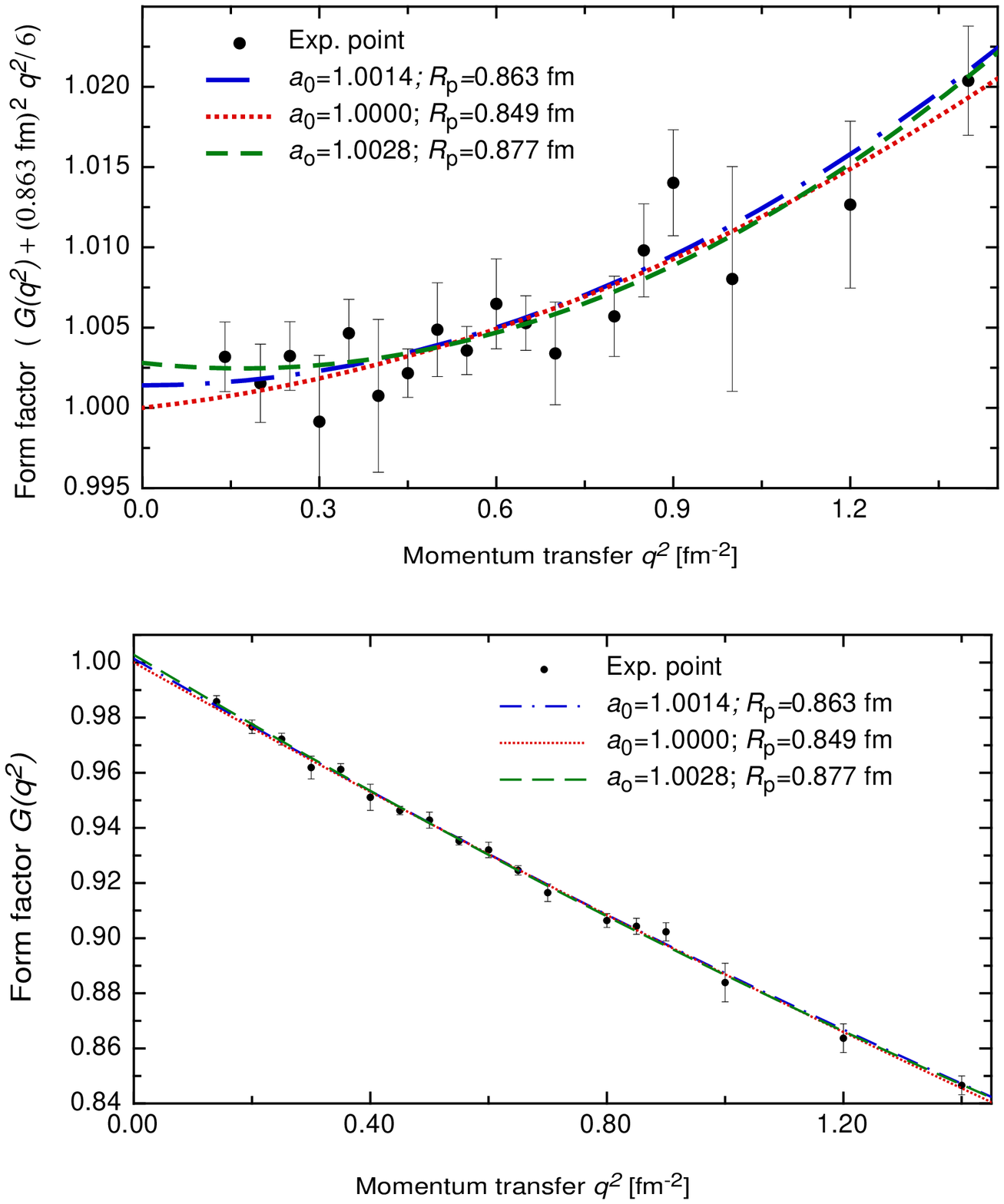}}
\caption{\label{f4} The Wong fits \protect{\cite{Wong}} of the Mainz-1980 data
\protect{\cite{Simon}}}
\end{figure}
We think that only the last fitting can be correct, because the normalization
constant $a_0$ cannot be measured with sufficient accuracy and has to be
determined by fitting. It has to be mentioned here that in Ref.
\cite{Simon} a special run of measurements
for the absolute gauging has been performed. The run
included 5 points with the systematic errors
equal to either 0.44\% or 0.46\% and the
statistical errors in a range between 0.42\% and 1.28\%. That means  that the last
fitting is actually in agreement with the experimental gauging within its
0.7\%-uncertainty and there is no reason to fix the normalization
as it was done in the first two fits (and in Refs. \cite{Simon,Mergell}.

The final uncertainty of the last result is
significantly larger than of the two first.
Such large changes of the uncertainty are due to two simple reasons. The
fit is almost a straight line. To determine it one has to know a few points from
the both
edges of the data. For the straight line the value of the reference interval
is quite important for the uncertainty.
When $a_0$ is fixed exactly one extra point at ${\bf q}^2=0$ is actually
included. The reference interval is larger, because for
the free normalization
one has to use actual results of the cross-section measurement and in this case
${\bf q}^2_{min}=0.14\; {\rm fm}^{-2}$. So we can see that for the free normalization
the reference interval is now smaller and the number of the fit parameters is
larger.

\section{Conclusions and discussions}

We can now summarize some results on the determination of the proton radius. To
the values presented in Fig. 1 we have added in Fig. 5 five other values. First we
have included without any discussion an old result with a more reasonable
uncertainty (that is fitting in Ref. \cite{Murphy} of the Saskatoon data). Next we
present in Fig. 5 the Wong fitting result with the free normalization
approach. We also give three values based one the hydrogen Lamb shift study.
One appears from using of the grand average value, and the others come from
the absolute measurements. Due to the discrepancy mentioned above between the
two Paris measurements we give the results for both $2s-8s/d$ transition
frequency values.

\subsection{What has been actually done}

\begin{figure}[h]
\epsfxsize=12cm
\centerline{\epsfbox{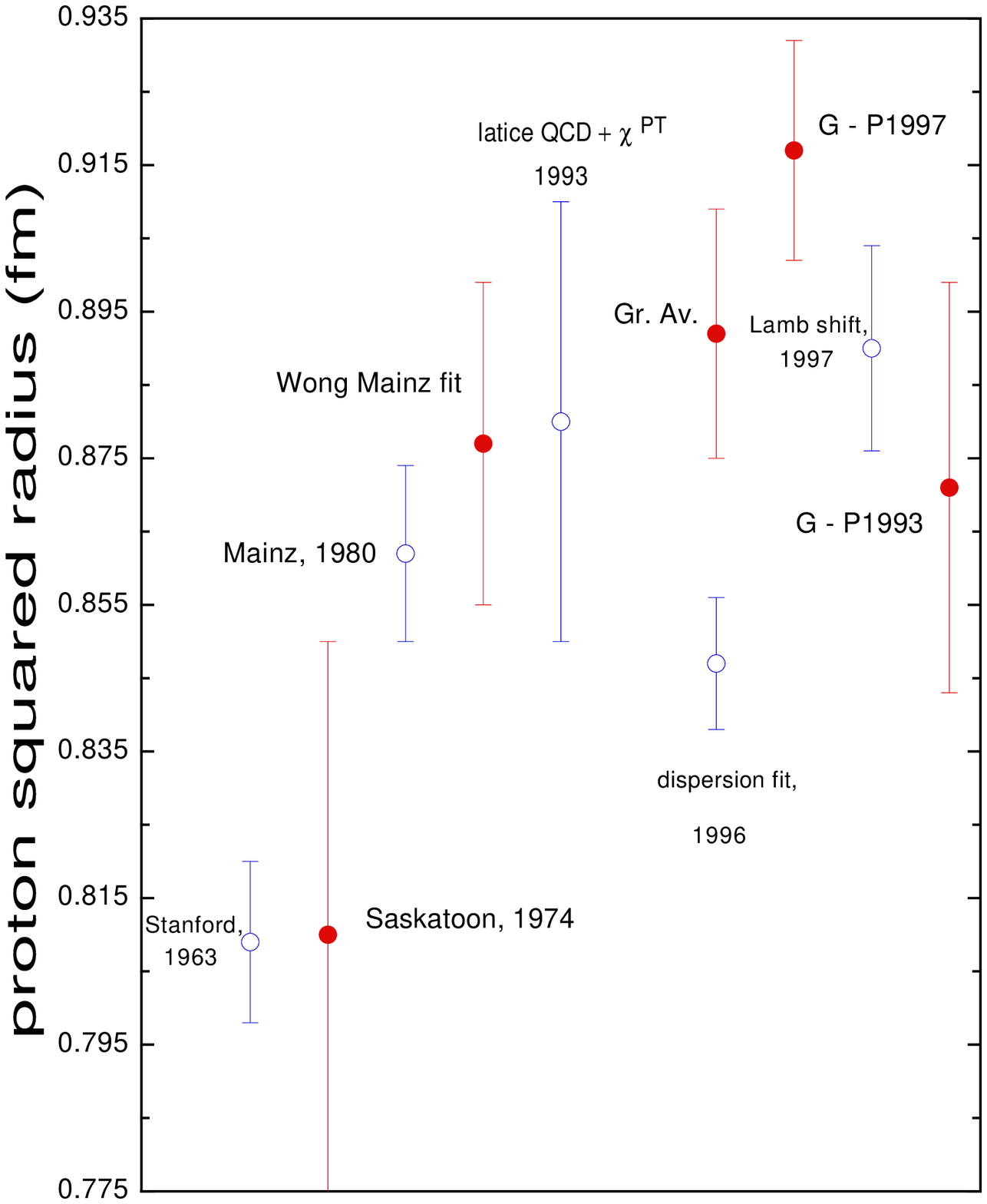}}
\caption{\label{f5} The different values of the proton radius as it has
to be}
\end{figure}

\begin{itemize}
\item The Lamb shift value for the proton charge radius significantly depends
on the data used. It is necessary to adjust all data. The uncertainty
includes 0.010 {\rm fm} (theory of the 1s Lamb shift), 0.002 {\rm fm} (theory of
$\Delta(2)$), 0.010 {\rm fm} (expected experimental uncertainty). The final
uncertainty is going to be 0.014 {\rm fm}.
\item The lattice calculations cannot give any good result at least because
only a few of the points for the form factor are found at moderate momenta. It
is not enough for a successful fitting. We also have some doubts due to the
uncertainty given in Refs. \cite{Draper,Leinweber93}.
\item The scattering data include an underestimation of the uncertainties.
The best
ones are from Mainz, and due to the Wong fitting with the free normalization it
seems that the reasonable uncertainty has to be close to 0.024 {\rm fm} (i. e. twice larger
than in the original work \cite{Simon}).
\item It is not clear what is the actual uncertainty of the radius
from the Mainz dispersion fitting and which part of data have actually
contributed to the result.
\end{itemize}

\subsection{What can be  done}

\begin{itemize}
\item One of the most important values for atomic applications is
$R_E^2-R_M^2$. Maybe it is less dependent on the proton  model.
That problem has
to be investigated. The value is to be helpful for extrapolating the low
energy data, for extracting $R_M$ from the Lamb shift measurement and
for calculating of the Zemach correction (see Appendix for details).
\item Adjustment for the overall hydrogen data has to include also results for the
proton radius. But the problem is: all results are correlated because
obtained by fitting of the experimental data. The newer compilations of them
are wider, but include also the older results.
\item Maybe a measurement of the muonic hydrogen Lamb shift can give a
more safe value, but that is an open question.
In case of the hydrogen-like ion of the helium-4, a high
accuracy has been obtained in an  experiment
\cite{Bertin}, but the result
has not been able to be reproduced by an other independent team
\cite{Hauser}.
So the safety can be just a problem.
\item It is not actually clear what the true uncertainties of
the Zemach correction with the Mainz dispersion fitting are.
\end{itemize}

\bigskip

This work is based on a talk given in September of
at the seminar at the Max-Planck-Institut f\"{u}r Quantenoptik and
the author would like to thank Prof. T. W. H\"ansch and all of the hydrogen
team at the Max-Planck-Institut f\"ur Quantenoptik for their support and
hospitality which make this work possible and for their interest to the problem
and numerous discussions which make the work more pleasant and
exciting. Malcolm Boshier is gratefully acknowledged for doing his
ZICAP transparencies available for us. The
support in part from the Russian state program `Fundamental Metrology' is also
acknowledged.

\appendix

\section{Hydrogen hyperfine structure and the proton structure}

The hyperfine splitting is much more sensitive to the proton structure
than the Lamb shift. The main nuclear
structure dependent contribution (so-called `Zemach correction') is of the form
\[
\Delta \nu(Zemach) = \nu_F\, \frac{2Z\alpha m_e}{\pi^2}
\int{\frac{d^3{\bf q}}
{{\bf q}^4}}\left[\frac{G_E(-{\bf q}^2)G_M(-{\bf q}^2)}{1+\kappa}-1\right]
\;.\]
The Zemach correction includes integration of the form factors,
but the most important part comes from the low momenta. That is because
the high momenta asymptotic behavior of the integral is determined by `-1'
and it is form factor independent. The low momenta asymptotics are defined by
eqs. (\ref{EAs}) and (\ref{MAs}) and the result for the Zemach integral
is proportional to $R_E^2+R_M^2$. So, the result depends straightly
on the both proton radius. The low momentum part of the integral
comes from the momentum
below $0.35\,{\rm Gev}\simeq 3\,{\rm fm}^{-2}$ and one can see from Figs. 3 and 4,
that in this region the approximation of \eq{EAs}
can give an appropriate expression for at least the electric form factor.

Some higher order structure dependent corrections were calculated by us
in Ref.
\cite{Karshenboim97c}, and all old results were also reviewed there.
That have allowed to give some prediction.
According to a classical review by {\sl G. T. Bodwin and D. R. Yennie, 1988}
\cite{Bodwin}, the comparison of theory and experiment leads
for the hydrogen hyperfine splitting to
\[
\frac{\nu_{hfs}(exp)-\nu_{hfs}(theo)}{\nu_{hfs}(exp)}=
\Big(0.48  \pm 0.56 \Big)\,ppm.
\]
\noindent
Proton polarizability is not included in $\nu_{hfs}(theo) $ and the
difference above has to be interpreted as its contribution.
The theoretical
limitation for the proton polarizability contribution is \cite{Hughes}
\[
\left\vert
\frac{\nu_{hfs}(exp)-\nu_{hfs}(theo)}{\nu_{hfs}(exp)}
\right\vert<4\,ppm.
\]
That is obtained from a treatment of the inelastic scattering data.
Our result for the comparison was \cite{Karshenboim97c}
\[
\frac{\nu_{hfs}(exp)-\nu_{hfs}(theo)}{\nu_{hfs}(exp)}=
\Big(3.4  \pm 0.9\Big)\,ppm.
\]
Due to the status of the magnetic radius which is not good determined
we need to say here that the uncertainty can be larger by factor like 3 from the
pure extrapolation data in the experimental work \cite{Simon}. Using the
dispersion fitting data \cite{Mergell} the result has to be shifted slightly,
but the
uncertainty has to be approximately the same (0.9 ppm), because
$R_E=0.847(9)\;{\rm fm}$ and $R_M=0.853(9)\;{\rm fm}$ and so the radii are equal one to
the other within their uncertainty. According the Wong fit the result for
the polarizability is rather close to 4(2) ppm within assumption,
that the magnetic radius is equal to the electric one.

\end{document}